\documentclass[twocolumn, amsmath, superscriptaddress, amsfonts,prb]{revtex4-2}

\usepackage{graphicx}
\usepackage{lipsum}

\pdfoutput=1

\begin{document}

\title{Voltage deficit in solar cells with suppressed recombination}

\author{Victor Karpov}\email{victor.karpov@utoledo.edu}\affiliation{Department of Physics and Astronomy, University of Toledo, Toledo,OH 43606, USA}
\author{Diana Shvydka}\email{Diana.Shvydka@osumc.edu}\affiliation{Department of Radiation Oncology, Ohio State University, Columbus, OH 43210, USA}

\begin{abstract}

The observed open circuit voltages  in best performing solar cells are explained outside of the recombination paradigm, based on such factors as electrostatic screening, Meyer - Neldel effect, and lateral nonuniformities. The underlying concept of suppressed recombination presents a long neglected alternative pathway to efficient PV. The criterion of suppressed recombination is consistent with the data for best performing solar cells. Also, consistent with the observations, is the open circuit voltage deficit that exhibits a lower bound of about $0.2-0.3$ V, does not correlate well with the optical gap, and shows a significant dispersion for materials possessing the same gap values. 
\end{abstract}

\maketitle

\section{Introduction}\label{sec:obs}
The many year quest for photovoltaic (PV) efficiency brought in significant improvements on many fronts. \cite{NREL2022,green2022} The understanding of PV operations evolved as well through improved modeling and new concepts. \cite{fahrenbrugh1983,handbook,nelson2003,wurfel2005,kar2021}  Through all the underlying developments, one  paradigm remaining intact was that of defect mediated nonradiative recombination [commonly termed Shockley-Read-Hall (SRH) \cite{sr1952,hall1952,sah1957,shockley1961}] dominating PV efficiency. \cite{fahrenbrugh1983,handbook,nelson2003,wurfel2005} Indeed, we are aware of just one publication addressing an alternative limitation due to the charge carriers extraction. \cite{bartesaghi2015}

A recent push towards an alternative explanation of the efficiency limitations was related to the case of PV with suppressed recombination \cite{efros2022} and based on the following approach. (i) Assume that there is no recombination, i. e. the rate of electron-hole pair generation equals the number of photons absorbed per time. In other words, we assume zero photocurrent loss. (ii) Extract the optimum power voltage $V_{mp}$ of a PV cell from its current-voltage characteristic, i. e. $V_{mp}$ is treated as an empirical parameter. The product $JV_{mp}$ presents the device power, which yields the efficiency.
How good is the above assumption of zero recombination in electric current?  The comparison of best cell efficiencies \cite{NREL2022,green2022} with calculations \cite{efros2022} validate that assumption to high accuracy.

Given a negligible effect of recombination on current, it is natural to expect its similarly weak effect on device voltage thus questioning the role of SRH recombination altogether.  There is however an alternative hypothesis claiming a significant deficit (loss) of PV voltages and relating it to recombination either directly or defined vaguely in terms of `material quality' exhibiting itself e. g. in Urbach tails \cite{chantana2020,chuang2015,kim2016,yi2020,park2018,shukla2021,wang2022,subedi2022,sood2022,koopmans2022,chen2020,kuciauskas2022,daboczi2019,halam2018}. More specifically, the open circuit voltage ($V_{oc}$) deficit is defined as
\begin{equation}\label{eq:voldef}\delta V_{oc}=G/e-V_{oc}.\end{equation}
The hypothesis behind the definition of Eq. (\ref{eq:voldef}) is that $V_{oc}$ is limited to the built-in voltage $V_{bi}$, which, in turn, is bound to the optical gap $G$. 

\begin{figure*}[t!]
\centering
\includegraphics[width=0.65\textwidth]{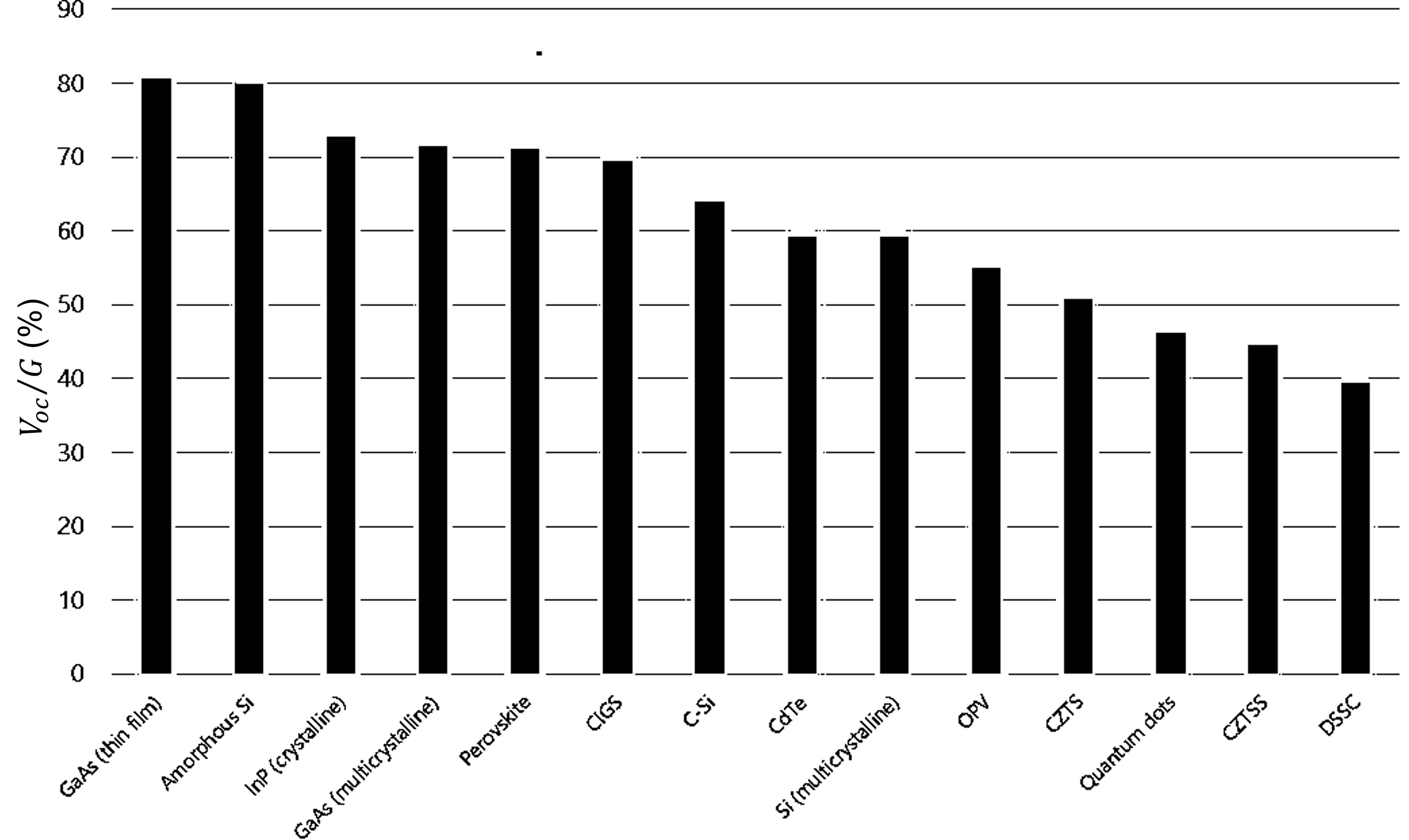}
\caption{The ratio $eV_{oc}/G\times 100 \%$ between the open circuit voltage $V_{oc}$ and optical gap $G$ for various PV brands; data from Ref. \cite{ossila}.
\label{Fig:VocG_orig}}
\end{figure*}

\begin{figure}[h!]
\centering
\includegraphics[width=0.35\textwidth]{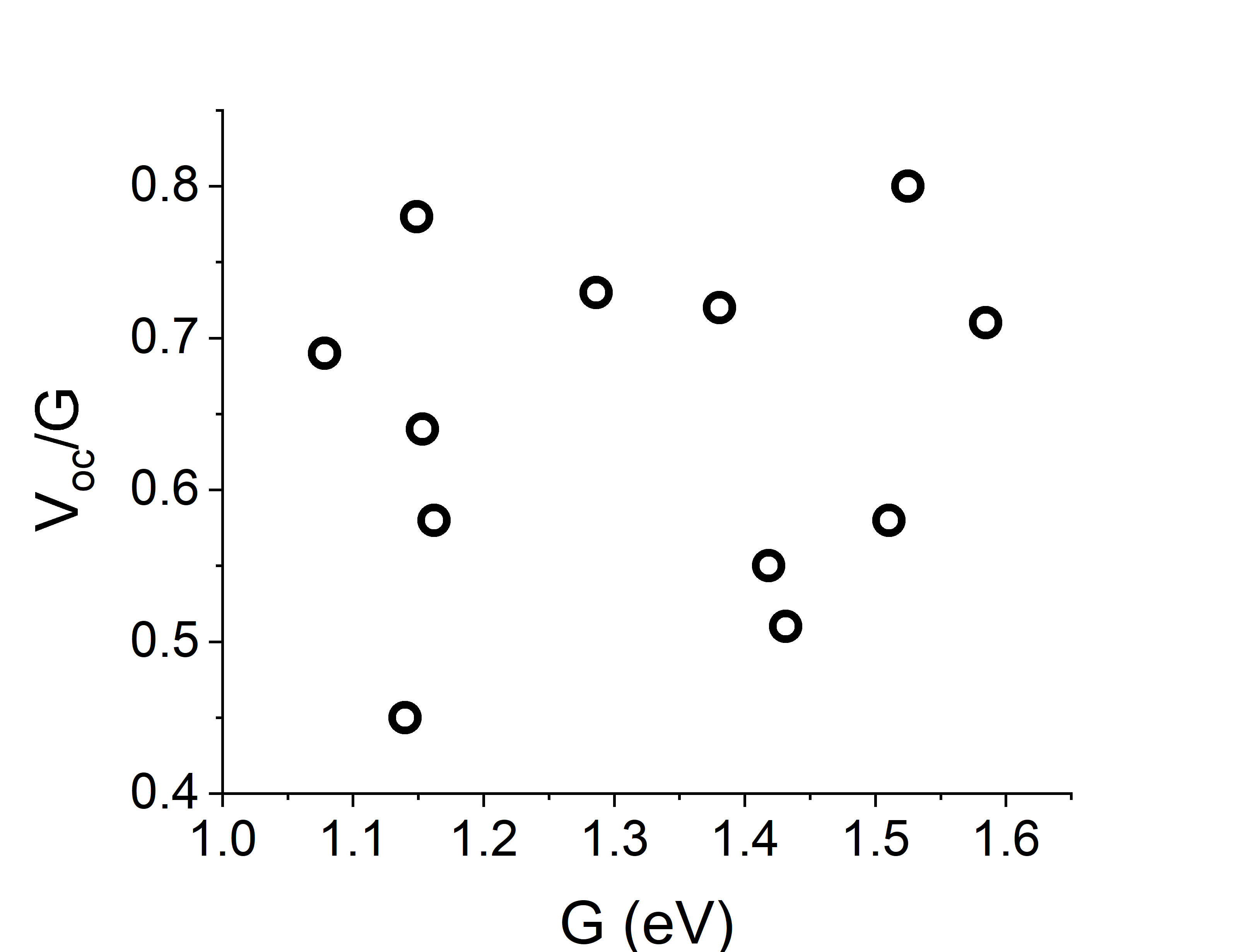}
\caption{The same data as in Fig. \ref{Fig:VocG_orig} after the materials are assigned their corresponding gap values from Table \ref{tab:VocG}. We have eliminated the data point for dye-sensitized PV as irrelevant to the material forbidden gap.
\label{Fig:VocG2}}
\end{figure}

The available data in Fig. \ref{Fig:VocG_orig} shows, at the first glance, a certain correlation, until we recognize that the horizontal axis is not quantified. Assigning the gap values to the materials leads to the data summarized in Table \ref{tab:VocG} and in Fig. \ref{Fig:VocG2}, which totally destroy the correlation.


\begin{table}[h!]
\caption{Open circuit voltages ($V_{oc}$) and optical gaps ($G$) for the main PV brands}
\begin{tabular}{|c|c|c|c|c|c|}
  \hline
  {\bf Brand}  & $V_{oc}$ (V) & G (eV) & {\bf Brand} & $V_{oc}$ (V) & $G$ (eV)\\ \hline
   GaAs TF  & 1.22 & 1.53&  c-Si& 0.738 & 1.115\\ \hline
   a-Si & 0.896 & 1.14 & CdTe& 0.876 & 1.51\\   \hline
 InP & 0.939 & 1.286 & Si mc & 0.674 & 1.16 \\   \hline
  GaAs - mc & 0.994 & 1.38 & OPV& 0.78 & 1.42 \\   \hline
   Perovs & 1.125 & 1.585 & CZTS& 0.73 & 1.43 \\   \hline
    CIGS & 0.744 & 1.078 & CZTSS& 0.513 & 1.14 \\   \hline
\end{tabular}\label{tab:VocG}
\end{table}

\begin{figure}[b]
\includegraphics[width=0.35\textwidth]{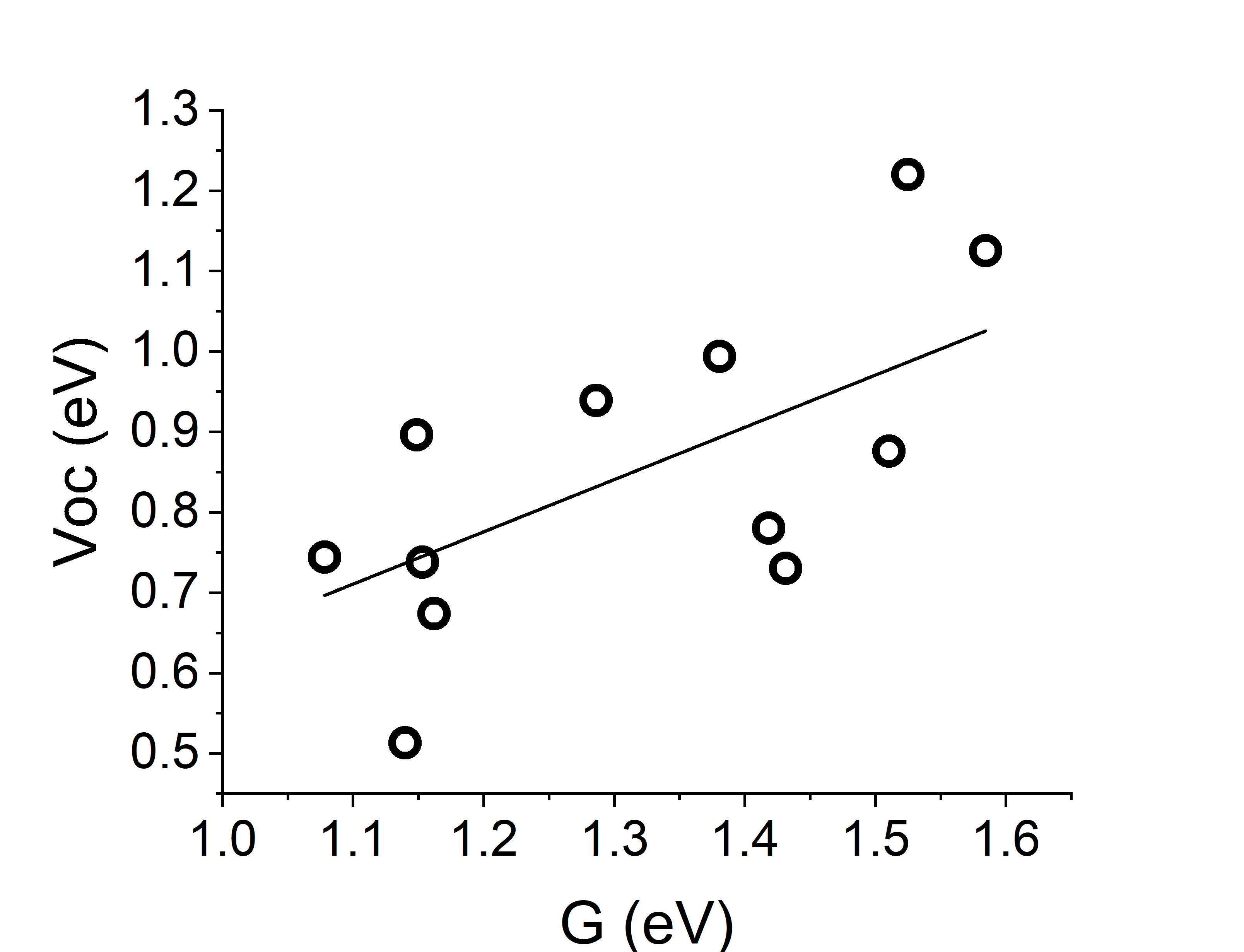}
\caption{Data for the best cells of various PV brands plotted following the input of Fig. \ref{Fig:VocG_orig} and Table \ref{tab:VocG}. Similar to Fig. \ref{Fig:VocG2}, we have eliminated the data point for dye-sensitized PV. The straight line represent the best linear fit. 
\label{Fig:VocG}}
\end{figure}


To set aside the role of recombination, we consider then the case of PV with suppressed recombination. That study revealed several voltage  affecting factors not related to recombination and calling upon certain practical remedies of minimizing the voltage deficit and improving PV efficiency. Simultaneously we will derive the criterion of suppressed recombination showing that it holds well for at least best thin-film PV.

%
%
%
\section{Nonradiative recombination: sufficient vs necessary }\label{sec:evid}

The PV literature is swarming with recombination topics showing how recombination dominated losses form sufficient basis for PV understanding. The later sufficiency means that SRH recombination theory \cite{sr1952,hall1952,sah1957,shockley1961} with adjustable defect parameters enables one to satisfactory interpret the data. However, sufficiency alone does not prove the causal relationship between recombination and performance. A necessity must be established in parallel where one condition (recombination)  must be present in order for another one (performance) to occur. Based on our extensive literature search, the assertion of recombination necessity is rarely, if at all, addressed. Indeed, such assertion would entail a formidable task of showing how the observed performance cannot be understood without the assumption of recombination loss. 

Here, we proceed along an alternative logical pathway showing how the observed performance can be explained without the notion of recombination. Should the latter be the case, recombination will lack the significance of general necessity giving a way to the paradigm of suppressed recombination PV. That paradigm does not imply zero recombination, but rather a recombination that is not as dominant as previously believed. Our evidence against the general necessity of recombination in PV is as follows. 
\begin{enumerate}
\item The physical processes underlying the recombination and drift processes are quite different, \cite{kar2021,abakumov1991} and so are their corresponding characteristic times. Therefore, it is natural to expect, in general, strong inequalities between the recombination time $\tau _r$ and drift time $\tau _d$: either $\tau _r\gg \tau _d$ (recombination practically irrelevant: the carriers leave the sample sooner) or $\tau _r\ll \tau _d$ (recombination totally dominates the PV efficiency). The possibility of recombination just moderately affecting efficiency, say, as often assumed, by 1-10 relative percent, i. e. $\tau _r\sim (10-100) \tau _d$ would be a sheer coincidence. In particular, it is reasonable to expect that recombination is irrelevant for the best PV at least.
\item Experimental data covering a variety of semiconductors exhibit the characteristic recombination times $\tau _r$ in the domain from microseconds to milliseconds, \cite{gray2002,milns1973,schroeder1997,parola2014,khatavkar2018,abakumov1991,landsberg1991, rein2005} much longer than the characteristic drift times $\tau _d\sim L/\mu  {\cal E}\sim 0.1-10$ ns for thin film PV. \cite{kar2021} For high quality materials,such as crystalline Si,  the inequality $\tau _r\gg \tau _d$ can hold even when $\tau _d$ represents the diffusion rather than drift time and device thickness reaches 100 $\mu$m.
\item The above mentioned recent publication \cite{efros2022} showing that totally neglecting recombination loss in the electric current yields rather accurate predictions for the observed PV efficiencies. Note that the latter argument is limited to the {\it best} performing cells. Our analysis in what follows will inherit the same limitation. (It is obvious indeed that low quality cells can suffer of  strong recombination.)
\end{enumerate}

The suppressed recombination concept significantly simplifies PV description, eliminates the need in multiple adjustable parameters, and shifts attention from the defect chemistry to basic physics. Surprisingly, that concept has never been seriously considered in PV science; we will analyze its predictions concentrating on possible non-recombination explanations of voltage deficit. 

\section{A simple PV model}\label{sec:scm}

\begin{figure*}[t!]
\includegraphics[width=0.7\textwidth]{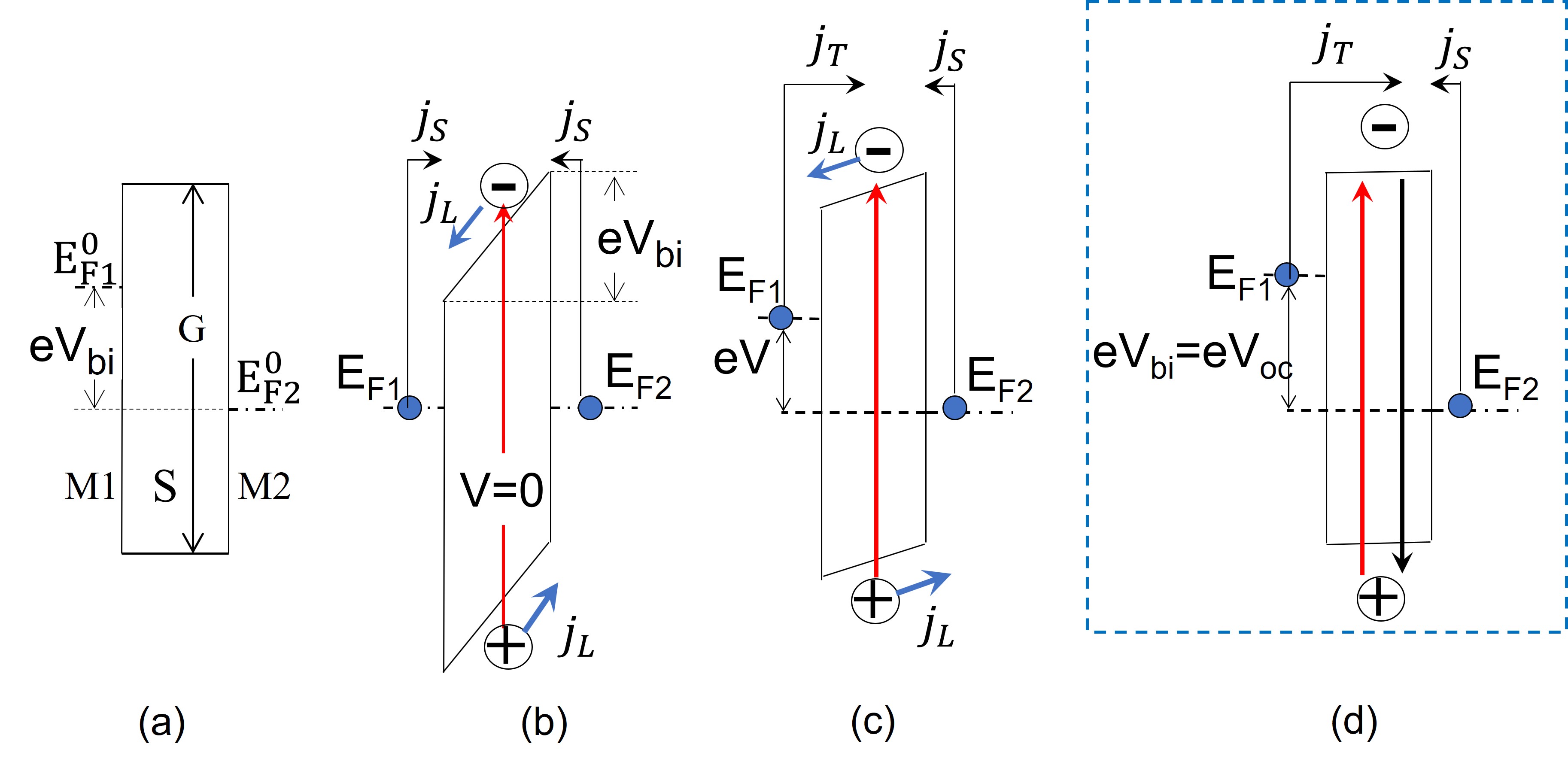}
\caption{Band diagrams of a cell formed by a semiconductor layer (S) with gap width $G$ sandwiched between two metals (M1, M2) with Fermi levels $E_{F1}^0$ and $E_{F2}^0$. Red vertical arrows represent photogeneration; the tilted blue arrows show the drift of charge carriers. (a) Bands before the electronic exchange between M1,S, and M2 is allowed. (b)The short-circuit condition with total voltage $V=0$ assuming zero voltage generation in a semiconductor layer. (c) Arbitrary voltage $V=[E_{F1}- E_{F2}]/e$, for different quasi-Fermi levels $E_{F1}$ and $E_{F2}$. Note the built-in voltage $V_{bi}=[E_{F1}^0- E_{F2}^0]/e\neq V$. (d) A dashed frame marked, inconsistent interpretation of zero built-in field case where photogenerated electrons and holes totally recombine (shown by downward arrow), and the significant thermal current $j_T$ is incompatible with the open circuit condition $j_{\rm total}=0$.
\label{Fig:bands}}
\end{figure*}

An ultimately simplified PV model in Fig. \ref{Fig:bands} accounts for the built-in uniform electric field between two metal electrodes.  The electrodes work function differential results in electric charge redistribution and its corresponding built-in electric field. 
Our model incorporates several simplifications:\\
(i) The layer is thin enough, so that the built-in field ${\cal E}$ is almost uniform.\\
(ii) A strong enough bias $V\gg kT/e$ when the diffusion component of current density can be neglected compared to that of drift, i. e. $j=env=en\mu {\cal E}$, where $n$ is the carrier concentration, $v$ is the velocity, $\mu$ is the mobility, and $kT$ has its standard meaning.\\
(iii) The mobilities $\mu$ and concentrations $n$ are about the same for electrons and holes.\\
(iv) There is no tunneling through the layer.\\
(v) The photogeneration is spatially uniform.\\
Relaxing the above simplifications will not significantly change our conclusions below. However, the following one is of principle importance:\\
(vi) We assume zero recombination rate. The corresponding criterion is derived  below.

For the case of suppressed recombination, the photocurrent $j_L$ (Fig. \ref{Fig:bands}) is voltage independent, corresponding to the condition that all photogenerated carriers are collected. (The possibility of voltage dependent photocurrent \cite{hegedus2007,zhao2021} would assume some recombination.) We note that in the approximation $j_L=const$, the charge carrier concentration $n$ will change with electric field ${\cal E}$ to maintain constant $j_L=n{\cal E}e\mu =g$ where $g$ is the carrier generation rate per area. 

In the standard current voltage characteristic,
\begin{equation}\label{eq:jV}j=j_S[\exp(eV/kT)-1]-j_L\end{equation}
$j_S$ is the saturation current. The open circuit regime $j=0$ corresponds to the balance of thermal and photo currents, \cite{fahrenbrugh1983,handbook,nelson2003, wurfel2005,kar2021}  $j_T-j_S=j_L$ where $j_T=j_S\exp(eV/kT)$. The physics behind such a balance is not the same as that of sometime assumed  scenario of `total recombination' between photogenerated electrons and holes under the condition of flat bands [Fig. \ref{Fig:bands} (d)] corresponding to zero electric field (M.J.Heben, private communication, November 03, 2022).

The balance $j_T-j_S=j_L$ along with Eq. (\ref{eq:jV}), yields the equation for open circuit voltage,
\begin{equation}\label{eq:Voc}V_{oc}=\frac{kT}{e}\ln\left(\frac{j_L}{j_S}+1\right)\approx \frac{kT}{e}\ln\left(\frac{j_L}{j_S}\right),\end{equation} showing how $V_{oc}$ is generally different from the built-in voltage $V_{bi}$ depicted in Figs. \ref{Fig:bands} (a), (b). We note for comparison that the above mentioned model of zero built-in field [Fig. \ref{Fig:bands} (d)] would nullify photocurrent $j_L$ at $V=V_{bi}\neq V_{oc}$, while leaving significant quasi-Fermi levels differential and its related forward current intact (we recall that the current is proportional to the gradient of quasi-Fermi levels; see e. g. Ref. \cite{koster2005}). Therefore, the open circuit condition is not directly related to $V_{bi}$. With that in mind, the very definition of voltage deficit in Eq. (\ref{eq:voldef}) as $V_{bi}-V_{oc}$ with $V_{bi}=G/e$ appears questionable. 

Note that using the gap dependent saturation current 
\begin{equation}j_{s}=j_{S0}\exp (-G/kT) \label{eq:satcur}\end{equation} 
where $j_{S0}$ is a material parameter, \cite{coutts2003} Eq. (\ref{eq:Voc}) predicts a  linear relation between $V_{oc}$ and $G$,
\begin{equation}\label{eq:Voc1}V_{oc}=G-\frac{kT}{e}\ln\left(\frac{j_{S0}}{j_L}\right).\end{equation}
Such a linear trend, while somewhat present in Fig. \ref{Fig:VocG}, is statistically weak, with the correlation coefficient of $\approx 0.38$ reflecting strong data dispersion.

We now compare the SRH related deterioration in electric current to that in voltage. The recombination can be described phenomenologically by modifying the carrier generation rate, $g\rightarrow g-R$ where $R$ is the number of electron-hole pairs recombining per area per time. The recombination will thus decrease the photocurrent by $\delta j_L=R$. For relatively small $R\ll j_L$ Eq. (\ref{eq:Voc}) predicts voltage degradation $\delta V_{oc} =kT\delta[\ln(j_L/j_S)]\approx V_{oc}R/g$. Therefore, the relative degradations in current and voltage are of the same order of magnitude,
\begin{equation}\label{eq:reldeg}\frac{\delta j_L}{j_L}\sim\frac{\delta V_{oc}}{V_{oc}}\sim \frac{R}{g},\end{equation}
i. e. SHR processes affect the current and voltage to approximately the same extent. Since in best devices the current is only slightly affected by recombination, so must be the voltage. We conclude again that SHR recombination may not be a significant factor in PV performance. 

\begin{figure}[h]
\centering
\includegraphics[width=0.35\textwidth]{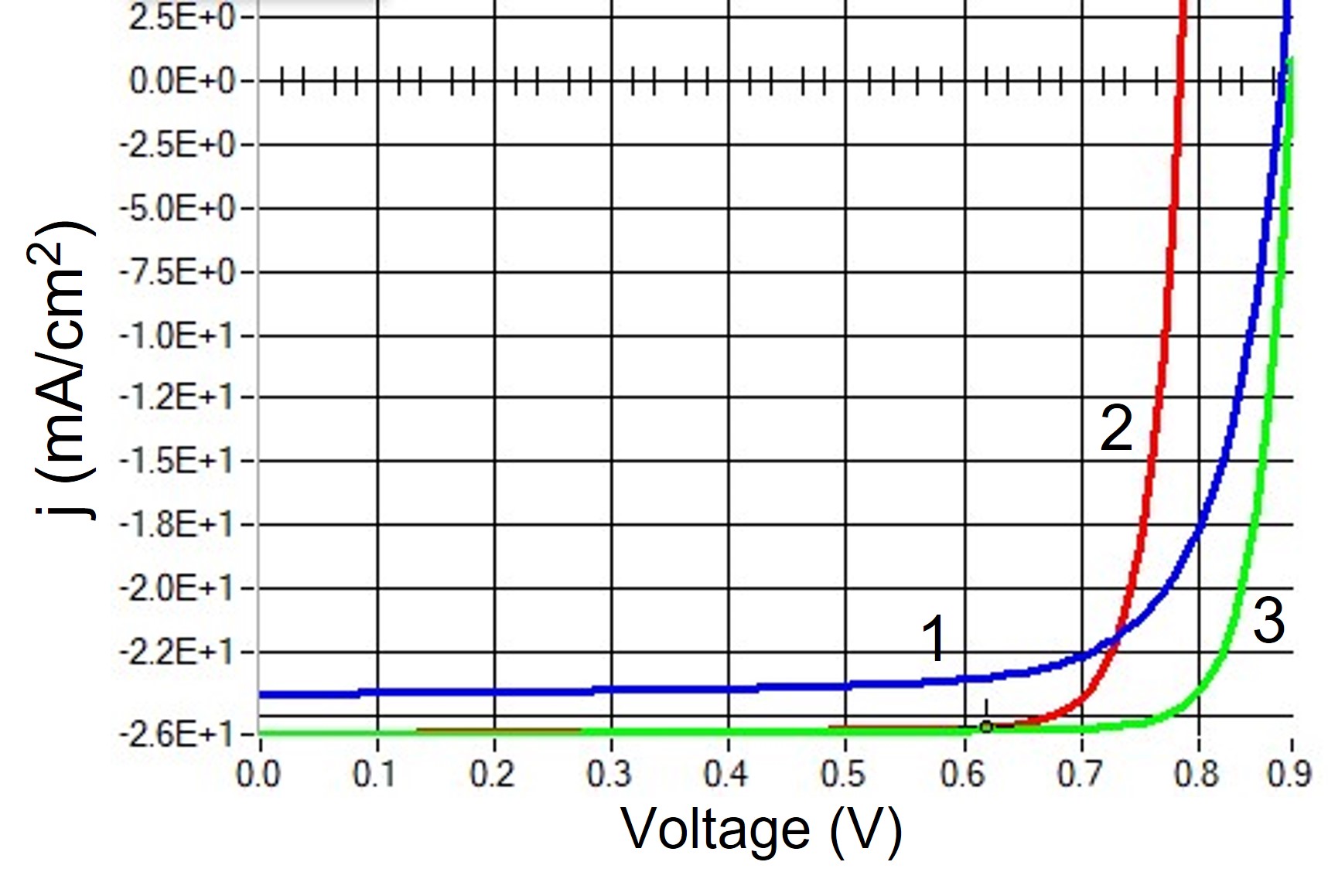}
\caption{SCAPS simulated current-voltage characteristics for the cases of 1 - Standard CdTe based PV \cite{gloeckler2003}, 2 - no defects in CdTe or CdS, but standard defect level in SnO$_x$, 3 -  no defects in all three components.
\label{Fig:IV}}
\end{figure}

For verification, we conducted SCAPS based numerical modeling for CdTe, CIGS and simple p-n systems, which together with the published perovskite results in Table \ref{tab:IV} confirm the same order of magnitude relative changes in current and voltage due to SHR recombination. Fig. \ref{Fig:IV} illustrates our approach to estimate the effect of recombination defect centers on CdTe based PV: starting with the `standard' model, we eliminated the recombination centers first from the semiconductor components, then from the TCO as well; we used a similar approach for CIGS and simple p-n junction models. For perovskites we used the results of published simulations for PV parameters vs. defect concentrations from two independent works. We would like to emphasize here that the role of those results is to show that relative effects on current and voltage are similar (rather than quantitatively predicting those effects). In particular, if the recombination effect on current is empirically found insufficient, one should expect the same for voltage.

\begin{table}[h]
\caption{Comparison of the defect caused relative current and voltage losses for various PV brands estimated through SCAPS modeling. The standard CdTe model \cite{gloeckler2003} was modified to eliminate its contained defects. Similar approach was applied to the CIGS and simple p-n models found with SCAPS distribution. \cite{SCAPS} Perovs1 and Perovs2 correspond to two independent cases of perovskite modeling, respectively \cite{du2016} and \cite{hussain2021} where the dependencies of defect concentration effects were explicitly determined.}
\begin{tabular}{|c|c|c|c|c|c|}
  \hline
  Loss  & CdTe & CIGS & simple p-n & Perovs1 & Perovs2 \\ \hline
   $\delta V_{oc}/V_{oc} \%$ & 9.5 & 2.5 &  16.9& 5.3 & 20 \\ \hline
   $\delta J_{sc}/J_{sc} \%$ & 6.6 & 5.5 & 7.8& 11.7 & 20.5 \\
  \hline
\end{tabular}\label{tab:IV}
\end{table}

\section{Criterion of suppressed recombination }\label{sec:polar}
When comparing the recombination time $\tau _r$ to the drift time $\tau _d=L/v$, a subtlety is that we do not immediately know the field strength ${\cal E}$. Both ${\cal E}$ and velocity $v=\mu {\cal E}$ can be affected by the redistribution of photogenerated carriers. Here we consider the possibility of such a redistribution effect.

\begin{figure}[t!]
\includegraphics[width=0.32\textwidth]{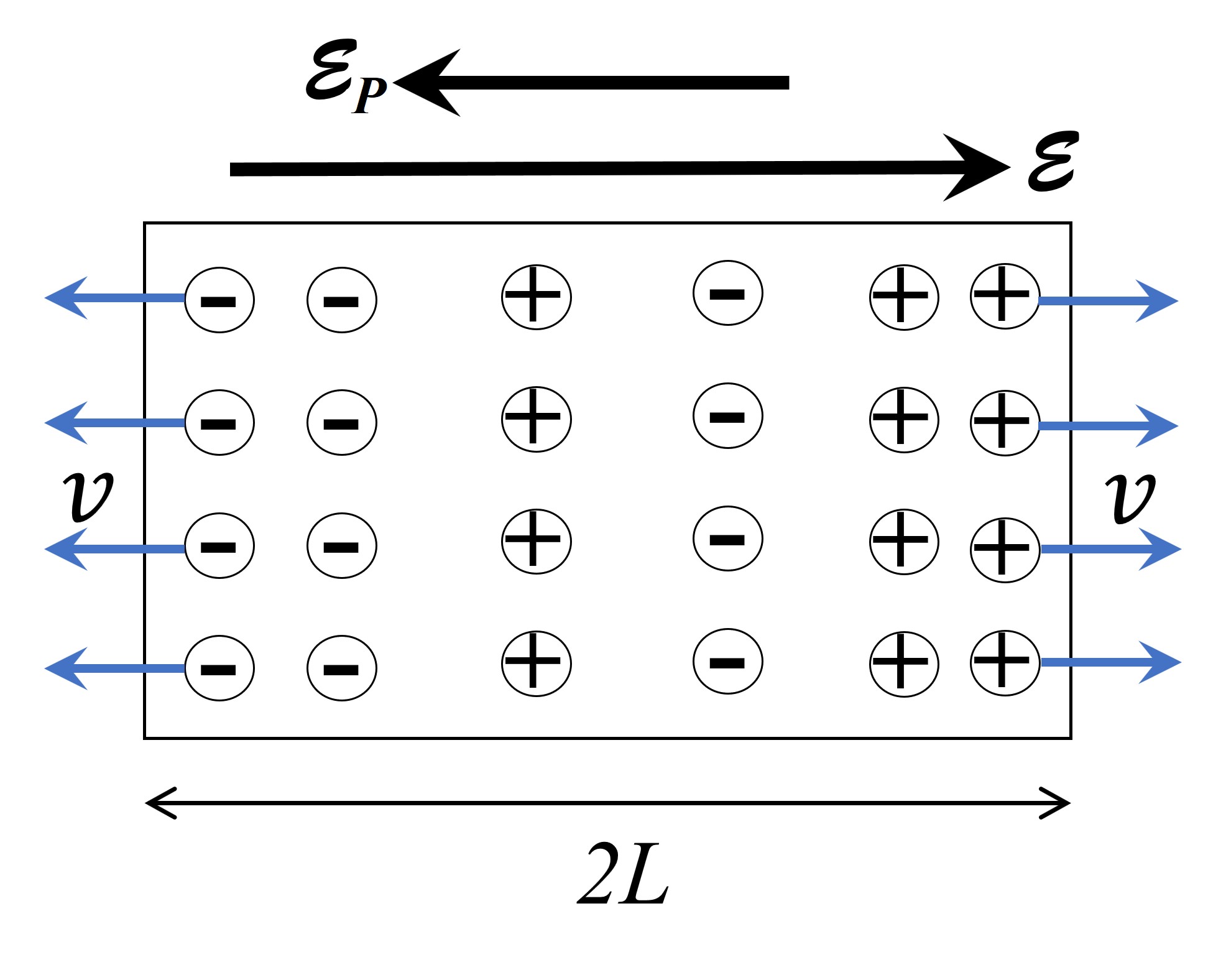}
\caption{A sketch of the free carriers charge distribution when the uniformly generated electrons and holes drift towards the opposite electrodes. ${\bf\cal E}$ and ${\bf\cal E_P}$ are respectively the original built-in field and polarization field.
\label{Fig:polar}}
\end{figure}

Sketched in Fig. \ref{Fig:polar}, are the photogenerated carriers moving towards the opposite electrodes and forming two oppositely charged layers of thicknesses $\sim L$ each. These layers create the polarization electric field ${\cal E}_P$ opposing the original built-in field ${\cal E}_{bi}=V_{bi}/2L$. The total electric field ${\cal E}_{bi}-{\cal E}_P$ is related to the electric potential difference $V_{bi}-V_P$  between the electrodes. Here we have introduced the photovoltage,
\begin{equation}\label{eq:deltaV} V_P={\cal E}_P2L= \alpha V_{bi}\end{equation}
with a heuristically defined parameter $\alpha \equiv{\cal E}_P/{\cal E}_{bi}$.

The absolute value of charge density per area is approximated as
\begin{equation}\label{eq:sigma}\sigma =ne =g\tau _de =g\frac{L}{\mu ({\cal E}_{bi}-{\cal E}_P)}e.\end{equation}
Here, $n=g\tau _d$ is the concentration of photogenerated carriers, $g$ is the generation rate (per area per time),
\begin{equation}\label{eq:tau}\tau _d=\frac{L}{\mu ({\cal E}_{bi}-{\cal E}_P)}\end{equation}
is the characteristic drift time accounting for the polarization effect.

The polarization field ${\cal E}_P$ becomes,
\begin{equation}\label{eq:EP}{\cal E}_P=\frac{4\pi\sigma}{\varepsilon}\end{equation}
(in Gaussian units) where $\varepsilon$ is the dielectric permittivity and $\sigma = nLe$ is the polarization surface charge density. Combining (\ref{eq:sigma}) and (\ref{eq:EP}) yields a quadratic equation,
\begin{equation}\label{eq:alpha}\left(\frac{{\cal E}_P}{{\cal E}_{bi}}\right)^2-\frac{{\cal E}_P}{{\cal E}_{bi}} +\alpha =0\end{equation}
with
\begin{equation}\label{eq:alpha1}\alpha \equiv \frac{4\pi gLe}{\varepsilon{\cal E}_{bi}^2\mu}=\frac{4\pi gL^3e}{\varepsilon V_{bi}^2\mu}.\end{equation}
It is natural to call $\alpha$ extraction delay parameter; its small values correspond to a favorable case of promptly extracted charge carriers.

To evaluate $\alpha$, we use with Eq.(\ref{eq:alpha1}) a set of realistic parameters, $4\pi/\varepsilon \sim 1$, $V_{bi}\sim 1$ V,\cite{sze,wiki} $L\sim 2-3$ $\mu$m, $\mu\sim 30$ cm$^2$V$^{-1}$s$^{-1}$, which yields $\alpha\sim 0.03-0.05$, i. e. $V_P\sim 0.03-0.05$ V. Small values of $\alpha \ll 1$ enable one to drop ${\cal E}_P$ in Eq. (\ref{eq:tau}), which reduces it to the intuitive estimate $\tau _d=L/\mu {\cal E}$.

\begin{figure}[t!]
\includegraphics[width=0.35\textwidth]{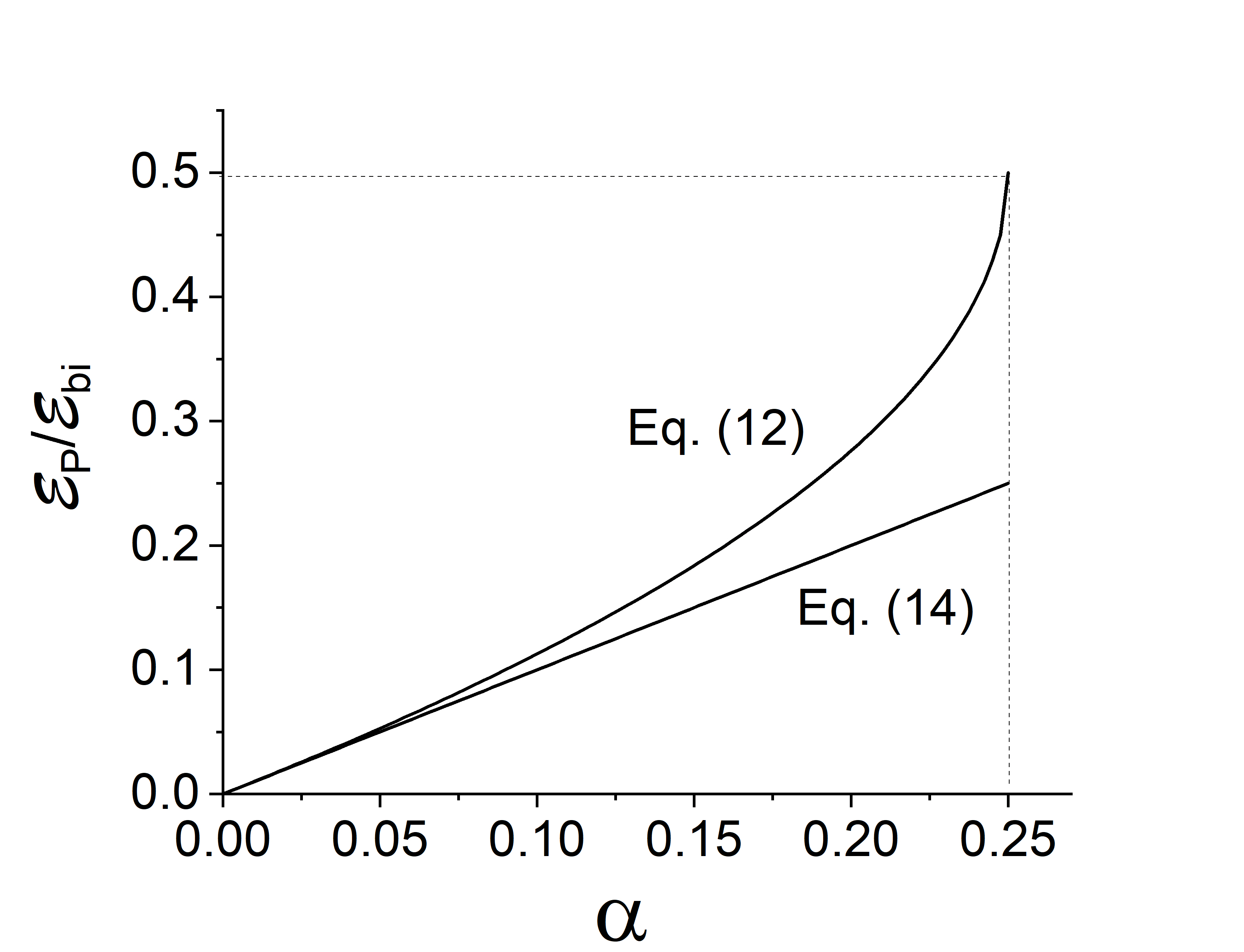}
\caption{The relative polarization field vs parameter $\alpha$ as predicted by Eqs. (\ref{eq:alpha}) and (\ref{eq:EPalpha}).
\label{Fig:alpha}}
\end{figure}

As follows from Eq. (\ref{eq:alpha}) and illustrated in Fig. \ref{Fig:alpha}, the photovoltage $V_P$ is limited to $V_{bi}/2$ when $\alpha = 0.25$. The upper bound ${\cal E}_P=V_{bi}/2$ reflects a maximum in polarization rate ${\cal E}_P/\tau _d$. For small $\alpha\ll 1$, Eq. (\ref{eq:alpha}) yields,
\begin{equation}\label{eq:EPalpha}{\cal E}_P=\alpha {\cal E}_{bi},\end{equation}
heuristically assumed in Eq. (\ref{eq:deltaV}). Also, our analysis predicts voltage independent photocurrent $j_L=2\sigma /\tau _d=2ge$ (the coefficient of 2 accounts for electrons and holes), consistent with the standard approach mentioned after Eq. (\ref{eq:jV}). 

On the other hand, for carrier mobilities several orders of magnitude lower, such as in polymer PV,  \cite{ebencoch2015,li2008} the parameter $\alpha$ can greatly exceed unity. For such low mobility structures, the drift time $\tau _d$ in Eq. (\ref{eq:tau}) becomes long enough to allow the dominant recombination; hence, the suppressed recombination concept does not apply. Accordingly, there is no reason to expect $j_L=const(V)$ in polymer PV, which conclusion is in agreement with empirical observations. \cite{koster2005,credgington2012}

In fact, the latter 'recombination sensitive' scenario can be analyzed similarly by replacing the drift time of Eq. (\ref{eq:tau}) with the composite $\tau _{eff}$ defined by,
\begin{equation}\label{eq:composite} \frac{1}{\tau _{eff}}=\frac{1}{\tau _d}+\frac{1}{\tau _r}.\end{equation}
including the recombination time $\tau _r$. As a result, Eq. (\ref{eq:alpha}) changes to the form,
\begin{equation}\label{eq:alphamod}\left(\frac{{\cal E}_P}{{\cal E}_{bi}}\right)^2-\left(1+\frac{L}{\tau _r\mu {\cal E}_{bi}}\right)\frac{{\cal E}_P}{{\cal E}_{bi}} +\alpha =0\end{equation}
where $\alpha$ is still defined by Eq. (\ref{eq:alpha1}). Along with $\alpha$, there is now another parameter describing the relative significance of recombination,
\begin{equation}\label{eq:beta}\beta\equiv \frac{L}{\tau _r\mu {\cal E}_{bi}}\end{equation}
We observe that the suppressed recombination criterion remains as suggested ($\beta\ll 1$), and large values of $\alpha$ are allowed when recombination matters, i. e. $\beta\gtrsim 1$. While it is straightforward to describe all possible cases of $\alpha$ and $\beta$, here we limit ourselves by concluding that the case of suppressed recombination and efficient PV corresponds to the limit $\alpha \ll 1,\quad \beta \ll 1$. 


\section{$V_{oc}$ for suppressed recombination} \label{sec:Voc}

Recombination out of the picture, very few possibilities remain to understand $V_{oc}$ data. As a general feature, these data in Fig. \ref{Fig:VocG} show (i) very significant dispersion of $V_{oc}$s, all well below the suspected ultimate values of $G$, and (ii) almost no correlation with $G$. A few possibilities underlying such a behavior and unrelated to SRH processes can be pointed out. \\

\begin{figure}[bht]
\includegraphics[width=0.2\textwidth]{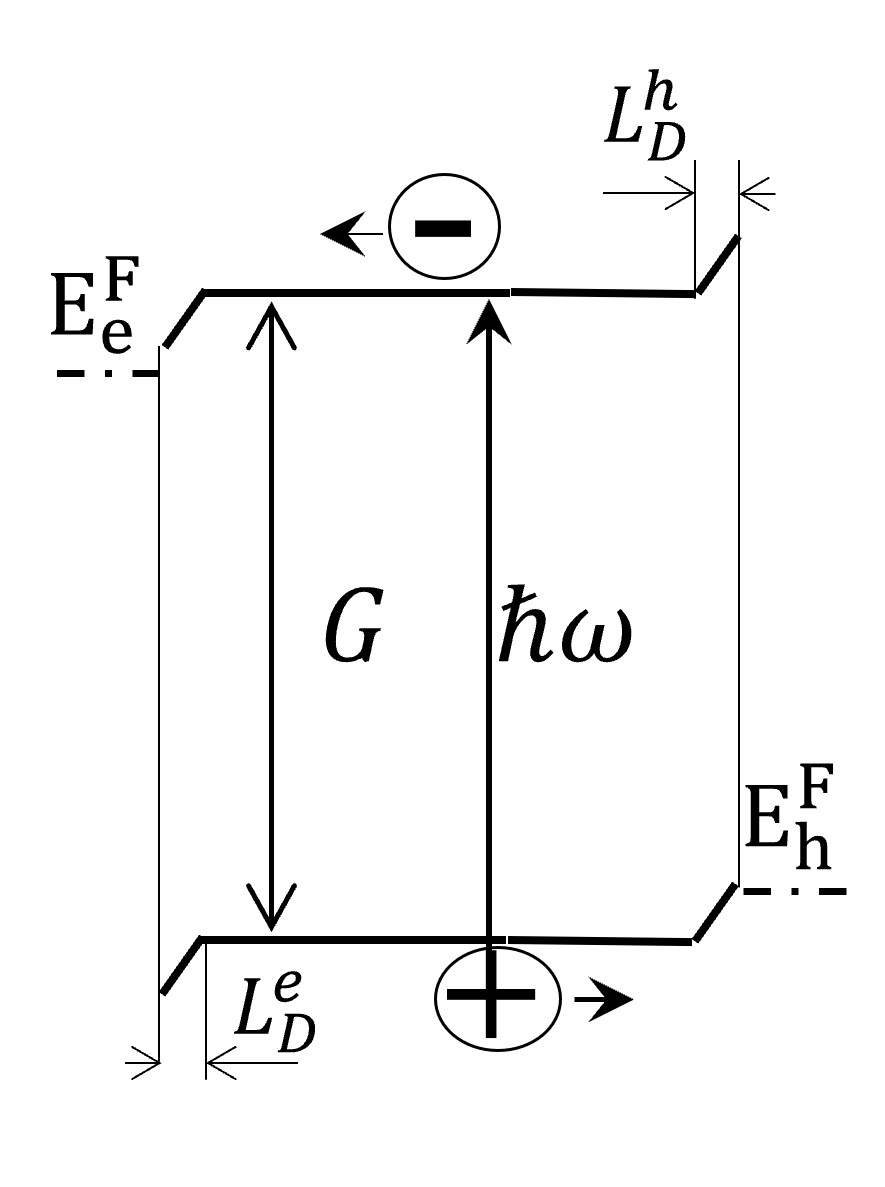}
\caption{The effect of electrostatic screening and field suppression when the quasi-Fermi levels are close to the corresponding band edges. Here $\hbar\omega$ is the absorbed photon energy. Note that the inner region of flat bands is qualitatively similar to the sketch in Fig. \ref{Fig:bands} (d), although they correspond to significantly different physical conditions.
\label{Fig:ulteffSQ}}
\end{figure}

1. {\it The electrostatic effect}. The desired zero voltage deficit assumes the difference between the electron and hole quasi-Fermi levels close to the corresponding conducting ($c$) and valence ($v$) band edges $E_c$ and $E_v$,
\begin{equation}\label{eq:QFL}E^F_e- E^F_h=eV_{oc}=G,\end{equation}
as illustrated in Fig.\ref{Fig:ulteffSQ}. Should the latter be the case, the concentrations of free charge carriers $n_{e(h)}=N_c^{e(h)}\exp(-|E_{c(v)}-E^F_{e(h)}|/kT)$ would become comparable to the effective state density, $N_c^{e(h)}=2(m^\star_{e,h} kT/2\pi\hbar ^2)^{3/2}$,  where $m^\star_{e,h}$ are the effective masses of electrons and holes. At room temperature $T=300$ K, one gets $N_c^{e(h)}\sim 10^{19}$ cm$^{-3}$. The high concentrations of charge carriers provide strong electrostatic screening, suppressing the electric field. Therefore, the photo-generated charge carriers will have to move by diffusion instead of drift, which significantly increases  their travel times, promoting recombination and degrading PV efficiency.

More quantitatively, the Debye screening lengths for electrons and holes are $L_D^{e,h}=\sqrt{\varepsilon kT/4\pi n_{e,h}e^2}$ where  $\varepsilon$ is the dielectric permittivity and $n_{e,h}$ is the corresponding charge carrier concentrations. Using $n_{e,h}\sim N_c^{e,h}\sim 10^{19}$ cm$^{-3}$ and $\varepsilon\sim 10$ one gets $L_D^{e,h}\sim 1$ nm, much thinner than the typical PV thickness $2L\sim 1$ $\mu$m. The electric field is not suppressed when  $L\leq L_D^{e,h}$. With Eq. (\ref{eq:QFL}) in mind, the latter inequality yields,
\begin{equation} \label{eq:criterion} \delta V_{oc}\geq \frac{kT}{e}\ln\left(\frac{L^216\pi N_c^{e,h}e^2}{\varepsilon kT}\right)\sim 10\frac{kT}{e}.\end{equation}
Here, the numerical coefficient of 10 is obtained assuming the typical numerical values of parameters involved. 

The criterion in Eq. (\ref{eq:criterion}) introduces the `ultimate' voltage deficit \cite{efros2022} controlled by the electrostatic screening. It is consistent with the collection of data in Fig. \ref{Fig:VocG2} and \ref{Fig:VocG} where none of the measured $V_{oc}$ is closer to $G$ than $\approx 0.25$ eV. Note that such `ultimate' voltage deficit does not depend on SRH recombination. While of the same order of magnitude for all conceivable semiconductors, it allows some moderate variations between them due to effective masses and working temperatures $T$.

The above estimate of ultimate voltage deficit can be affected by considering electrostatic screening by localized states \cite{mott1979,kar2021}, e. g. band tail states. We do not expect the corresponding revision to be very significant because the concentrations of tail states are typically comparable to $N_c^{e,h}$. Yet such screening modification can make the ultimatet deficit more sensitive to material structure.\\

2.{\it Meyer-Neldel (MN) rule} observed in a great variety of semiconductor structures. It states that the probabilities of thermally activated processes include non-thermal exponentials in the form,
\begin{equation}\nu = \nu _0\exp (-E/kT+E/kT_{MN})\label{eq:MNR}\end{equation}
where $T_{MN}$ is a fictitious system-sensitive temperature often in the interval of $T_{MN}\sim 600-1000$ K; in some cases, negative $T_{MN}$ of similar absolute values are reported. \cite{coutts1984,goradia1984,seymour2005,prasai2016,shmidt2007,yelon2006,starikov2014,widenhorn2004,takamure2022,savransky2010} The physics behind MN rule remains disputable with two popular hypothesis that (a) the entropy contribution $TS$ to the system free energy $E-TS$ is proportional to $E$ (`compensation rule') because of the exponentially large number of combinations of small energy ($\epsilon\ll E$) excitations, for example, phonons, involved, or (b) the activation barrier energies are random. Setting aside the question of its origin, we note here that the MN rule is not related to any type of electron-hole annihilation and thus is consistent with the concept of suppressed recombination. 

Taking the MN rule into account Eqs. (\ref{eq:voldef}) and (\ref{eq:Voc1}) combine to yield,
\begin{equation}\label{eq:Voc2}\delta V_{oc}=\frac{AkT}{e}\ln\left(\frac{j_{S0}}{j_L}\right), \quad  A\equiv \frac{T_{MN}}{T_{MN}-T},\end{equation}
which is qualitatively consistent with the above mentioned observations that $eV_{oc}$'s are considerably lower than $G$'s and rather weekly correlated witg $G$'s (since $T_{MN}$ is independent of $G$). Note that $A$ in Eq. (\ref{eq:Voc2}) links the MN rule to the diode ideality factor $A$ correctly predicting the typically observed $A>1$. 

While interpreting $T_{MN}$ as an inherent material parameter, the result in Eq. (\ref{eq:Voc2}) leaves the leakage current $j{S0}$ potentially attainable to technology. Furthermore, variations in $j{S0}$ can explain variations in open circuit voltage between nominally identical devices, observed more often than reported.\\

2. {\it Leakiness} of ohmic or non-ohmic nature not affecting the photocurrent, yet noticeably lowering $V_{oc}$ can be due to microelements known as week diodes. \cite{kar2021,karpov2002} Phenomenologically, the effect can be attributed to the local increase in the saturation current $j_{S0}$ determining $V_{oc}$ through Eq. (\ref{eq:Voc1}). 

\begin{figure}[bht]
\includegraphics[width=0.35\textwidth]{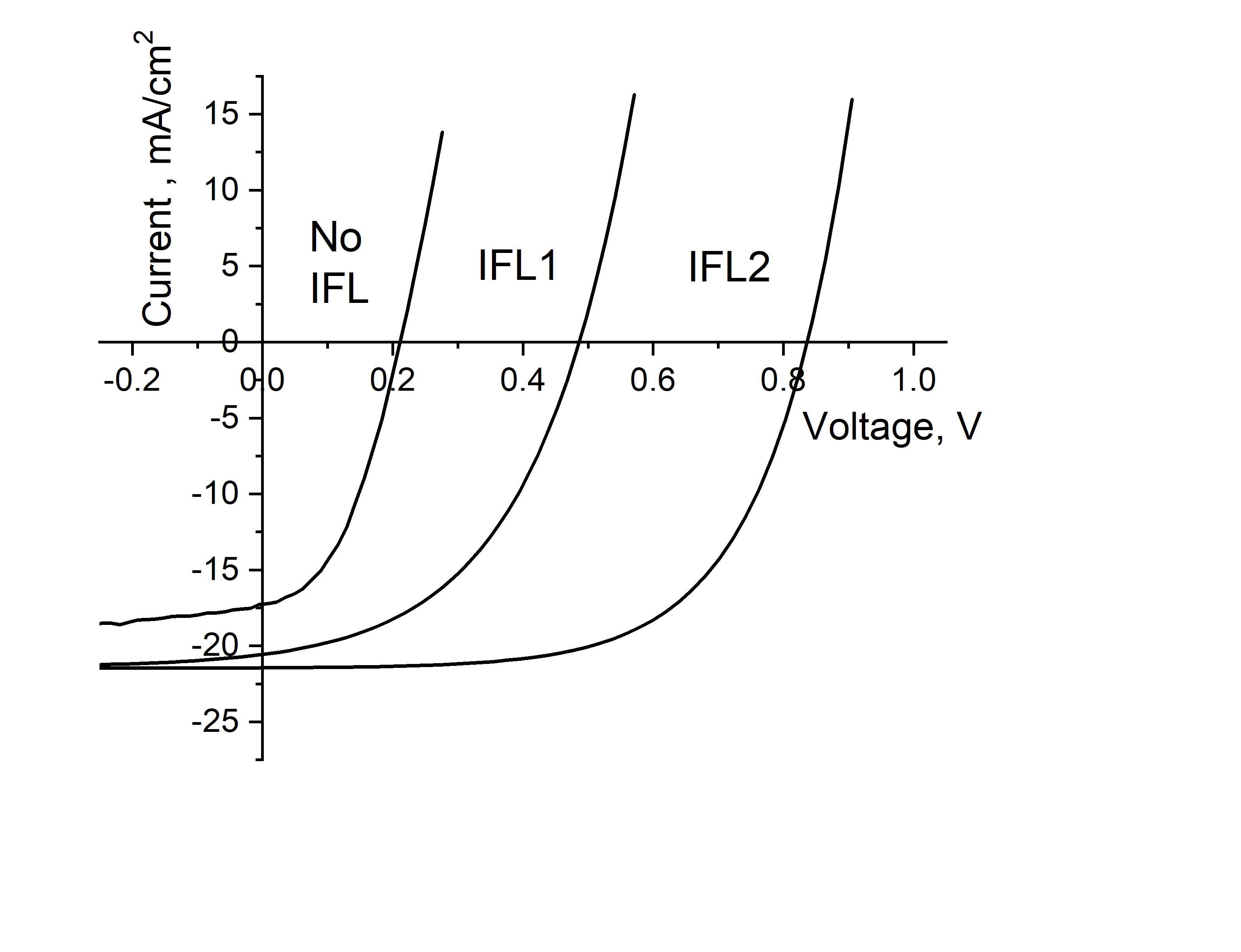}
\caption{The interface related effects in nominally identical PV CdTe based cells with different interfacial layers, for brevity denoted here as IFL1 and IFL2. Note very strong changes in $V_{oc}$ and relatively small changes in the short-circuit current. \cite{roussillon2004}
\label{Fig:IFL}}
\end{figure}

At a more microscopic level, the system is represented with a set of microdiodes possessing random local open circuit voltages $v_{oc}$ (and approximately the same photocurrents), which are connected in parallel through a resistive electrode. The microdiode lateral dimensions are on a micron scale, $\l\sim 1$ $\mu$m. Those possessing low $v_{oc}$s ({\it weak diodes}) are responsible for the macroscopic voltage deficit. The weak diodes are often related to interfacial properties. One illustration is presented in Fig. \ref{Fig:IFL} based on the previously published data. \cite{roussillon2004}.

A variety of microscopic mechanisms \cite{kar2021} can be responsible for the weak diode behavior, mostly implying leakage due to percolation in nonuniform films, such as hopping conduction through the localized states, possibly including the Urbach absorption related tail states. All of the weak diode models, perhaps even different between different materials, are reducible to the local spots of anomalously high suturation current density $j_S$ and $j_{S0}$ that translates to low microscopic $v_{oc}$ characterizing such spots. The presence of leaking spots was first observed decades ago in connection with the thermionic emission \cite{becker1935,herring1949} understood then in terms of the patch model accounting for lateral variations in metal work functions. Later, the concept of lateral variations was recognized for thin amorphous films \cite{pollak1973} and other disordered systems \cite{raikh1991,baranovskii1997} and for thin film PV. \cite{kar2021,karpov2002} It was demonstrated  recently that for the case of thin films, the leaking pathways are almost rectilinear.\cite{patmiou2020} 

For the open circuit regime, the condition of zero total current leads to the macroscopic (average) open circuit voltage
\begin{equation}\label{eq:Vocmacr}V_{oc}=-\frac{AkT}{e}\langle \exp\left(-\frac{ev_{oc}}{AkT}\right)\rangle \end{equation} 
where the angular brackets stand for averaging over random microscopic $v_{oc}$. Eq. (\ref{eq:Vocmacr}) explicitly shows how anomalously low (weak diode) $v_{oc}$ entities have exponentially strong effect on the observed $V_{oc}$. Physically blocking such weak spots with interfacial layers will increase $V_{oc}$. This is illustrated in Fig. \ref{Fig:IFL} where the existing weak diode spot is blocked by applying this or other interfacial layer (IFL). 

Averaging in Eq. (\ref{eq:Vocmacr}) becomes rather nontrivial \cite{kar2021,karpov2002} when multiple comparably weak diodes act simultaneously within the same electrically connected domain of size  
\begin{equation}\label{eq:L0}L_0=\sqrt{\frac{AkT}{e\rho j_L}}\end{equation}
where $\rho$ is the electrode sheet resistance. All subtleties  \cite{kar2021,karpov2002} aside, the approximate result for the weak diode caused voltage deficit is given by
\begin{equation}\label{eq:Voc3}\delta V_{oc}\approx\frac{2kT}{e}\ln\left(\frac{L_0}{l}\right)\end{equation}
where $l$ is the characteristic diameter of a weak diode. The ratio $L_0/l\gg 1$ can be as large as $10^3-10^4$ corresponding to $\delta V_{oc}$ of several tenths of one Volt in Eq. (\ref{eq:Voc3}). Similar to the MN related $\delta V{oc}$, Eq. (\ref{eq:Voc3}) explains the nature of $\delta V_{oc}$ and the data spread for materials possessing the same $G$. Furthermore, variations in microscopic morphology can explain the often observed dispersion in open circuit voltages between nominally identical devices. 

Under light, the weak diode entrances in the film interface form local spots of different electric potential, which drive lateral electric currents playing the role of current sinks. On the other hand, they can attract mobile ionic particles capable of neutralizing such spots with proper IFL. In particular, electrolytes or other passivation agents can mitigate the weak diode effects \cite{yi2020,wang2022,sood2022,kar2006}. Furthermore, if not treated, the weak diode effects can aggravate due to the positive feedback of persistent flow of localized leakage current degrading the cell performance. Therefore, passivation treatments can improve the device stability as well. 

The weak diode related voltage deficit can be experimentally identified by various mappings \cite{kar2021,karpov2002}, by size effects leading to stronger variations among smaller size devices, and by the light intensity dependent measurements when low light conditions increase $L_0$ and make voltages more dispersed. \cite{kar2021,karpov2002} \\

The above three sources of voltage deficit are all unrelated to SRH recombination, and yet they can explain the observed significant $\delta V_{oc}$s, the existence of `ultimate' $\delta V_{oc}\approx 0.2-0.3$ V and $\delta V_{oc}$s dispersion between devices possessing almost the same optical gaps. To avoid any misunderstanding we note that if $\delta V_{oc}$ predicted by Eqs. (\ref{eq:Voc2}) or Eq. (\ref{eq:Voc3}) exceeds that of Eq. (\ref{eq:criterion}), the latter should not be taken into account (in particular, no need to add it). 

A number of uncertainties (such as the electrostatic screening by tail states, the microscopic nature of weak diodes) and unknowns (e. g., the nature of Meyer-Neldel effect) make the comprehensive understanding of suppressed recombination PV to be work in progress.

\section{Conclusions}\label{sec:concl}
In summary, we have demonstrated the following: 
\begin{enumerate}
\item The known definition of voltage deficit [Eq. (\ref{eq:voldef})] may lack sufficient empirical and theoretical grounds as a relevant voltage parameter.
\item There exist the conditions of suppressed recombination and efficient carrier extraction, under which SRH recombination processes are insignificant. These conditions hold well for at least best performing thin film PV.
\item There are physical factors (electrostatic screening, MN effect, and leakage) not related to SRH recombination that can significantly affect the observed  PV voltage.
\item Some uncertainties and unknowns remain calling upon future work towards understanding the limitations of suppressed recombination devices.  
\end{enumerate} 


The suppressed-recombination physics presented here may not be in synergy with a culture of recombination dominated PV. \cite{scarpulla2023} The well known historical circumstances underlying the latter can explain its overwhelming acceptance: (a) the first semiconductor devices were relatively thick thus giving enough time for the electrons and holes to recombine before they reach the electrodes;  (b) the material quality of the first semiconductors was rather poor, with lots of imperfections promoting recombination. Hence, the SRH theory of recombination explaining the observations became a significant part of semiconductor physics.

The circumstances changed since: (a) semiconductor films of modernity are thin, down to sub-micron scales where electrons and holes may have a good chance to get extracted before they recombine; (b) the modern materials are cleaner, with fewer imperfections, so the SRH recombination may be not that strongly promoted (even in polycrystalline films structural imperfections are pushed away from the grains during crystallization; they end up at grain boundaries where their role may turn beneficial due to electric charging and coulomb repulsion forming recombination barriers \cite{kar2021} ). Therefore, we call for a careful  differentiation between the qualities of {\it sufficient vs necessary} for the paradigm of recombination driven PV. From the practical perspectives, adopting the concept of suppressed recombination will shift the research efforts from defect chemistry to interfacial treatments and more fundamental issues such as screening and Meyer - Neldel rule.


Finally, our results here are applicable to other devices call for other developments towards suppressed recombination devices, such as thin-film semiconductor photo-detectors or light-emitting diodes. We point at our recent successful attempt to understand X-ray perovskite detectors based on suppressed recombination paradigm. \cite{efros2022}

\end{document}